\documentclass[twocolumn,preprintnumbers,pra,smaller,
superscriptaddress]{revtex4}

\usepackage{amssymb,amsmath,graphicx,dcolumn,bm,float} 
\usepackage[utf8]{inputenc}
\usepackage[T1]{fontenc}
\usepackage[colorlinks,linkcolor=black,citecolor=blue]{hyperref}

\begin{document}

\title{Self-organized synchronization of mechanically coupled resonators based on optomechanics gain-loss balance}
\author{P. Djorwé}
\email{philippe.djorwe@icn2.cat}
\affiliation{Institut d’Electronique, de Microélectronique et Nanotechnologie, UMR
CNRS 8520 Université de Lille, Sciences et technologies, Villeneuve d’ Ascq 59652, France}

\affiliation{Catalan Institute of Nanoscience and Nanotechnology (ICN2), BIST, Campus UAB, Bellaterra, 08193 Barcelona, Spain}

\affiliation{Department of Physics, Faculty of Science, The University of Ngaoundere, P.O. Box 454, Ngaoundere, Cameroon}

\author{Y. Pennec}
\email{yan.pennec@univ-lille1.fr}
\affiliation{Institut d’Electronique, de Microélectronique et Nanotechnologie, UMR
CNRS 8520 Université de Lille, Sciences et technologies, Villeneuve d’ Ascq 59652, France}

\author{B. Djafari-Rouhani}
\email{bahram.djafari-rouhani@univ-lille1.fr}
\affiliation{Institut d’Electronique, de Microélectronique et Nanotechnologie, UMR
CNRS 8520 Université de Lille, Sciences et technologies, Villeneuve d’ Ascq 59652, France}

\begin{abstract}
We investigate self-organized synchronization in a blue-detuned optomechanical cavity that is mechanically coupled to an 
undriven mechanical resonator. By controlling the strength of the driving field, we engineer a mechanical gain that balances 
the losses of the undriven resonator. This gain-loss balance corresponds to the threshold where both coupled mechanical 
resonators enter simultaneously into self-sustained limit cycle oscillations regime. This leads to rich sets of collective dynamics 
such as in-phase and out-of-phase synchronizations, depending on the mechanical coupling rate, the frequency mismatch between the resonators, and the external driving strength through the mechanical gain and the optical spring effect. Moreover, we show that the introduction of a quadratic  coupling enhances the in-phase synchronization. This work shows how phonon transfer can optomechanically induce synchronization in a coupled mechanical resonator array and opens up new avenues for metrology, phonon-processing, and novel memories concepts.
\end{abstract}

\maketitle

\date{\today}


%
\section{Introduction} \label{Intro}

Recent progress in nano engineering has led to exotic optomechanical crystals \cite{Eichenfield2009, Eichenfield_2009, Chang.2011.NJP},  which are able 
to confine several optical and mechanical modes on a single chip. A lot of attention has been paid 
to these optomechanical arrays owing to their capabilities to promote new phenomena and applications. 
These include collective nonlinear dynamics \cite{Heinrich.2011},
\cite{Wallin.2018}, quantum many-body dynamics of photons and phonons \cite{Ludwig.2013.Sync}, long-range coupling of phonon 
modes \cite{Xuereb.2012, Xuereb.2014}, photons and phonons transport \cite{Chen.2014, Schmidt.2015}, Anderson localization \cite{Roque.2017}, as well as topological phases of sound and light \cite{Peano.2015.PRL, Brendel.2018}.

Owing to their  practical applications in rf communication \cite{Bregni.2002}, signal-processing \cite{strogatz2003}, and clock synchronization \cite{bahder2009}, collective phenomena such as synchronization 
\cite{Zhang.2012, Zhang_2015, Bagheri.2013, Colombano.2019, Sheng.2020} and frequency locking \cite{Shah.2015, Gil.2017} were recently realized in optomechanical systems. These schemes are different one from  another, and 
each of them having its own specificity.  Synchronization of two oscillators in contact
and sharing a common optical mode was investigated in \cite{Zhang.2012}, while the same configuration was later on extended to multiple resonators in \cite{Zhang_2015}. Another synchronized system, consisting of two spatially separated oscillators
integrated in a common optical racetrack cavity was reported in \cite{Bagheri.2013}. Synchronization of the mechanical dynamics of a pair of optomechanical crystal cavities, which are intercoupled with a mechanical link and support independent optical modes, has been reported in \cite{Colombano.2019}. The observation of anti-phase synchronous oscillations has been observed in this system, which is the first reported work on mechanically coupled resonators in optomechanics. More recently, self-organized synchronization of two optically coupled membranes placed inside an optomechanical cavity has been demonstrated in \cite{Sheng.2020}. It has been shown that the system enters into the synchronization regime via a torus birth bifurcation line. Also, the phase-locking phenomenon and the transition between in-phase and anti-phase regimes were directly observed. Besides of these optomechanical synchronization investigations, frequency locking of two and multiple optomechanical systems were reported respectively in \cite{Shah.2015} and \cite{Gil.2017}. In all these works, light is the key element used to couple the optomechanical systems involved except in \cite{Colombano.2019}. It follows that there is a lack of work on optomechanical synchronization where phonons are mediating the coupling between the oscillators. Theoretical investigation of quantum many-body dynamics has been investigated along this line in \cite{Ludwig.2013.Sync}, where both photons and phonons can hop to nearest neighbor sites, 
and each optomechanical cell in the array is driven. 

In the present work, we propose a blue-detuned optomechanical system that is mechanically coupled to an undriven 
mechanical resonator. By driving the optomechanical system, we engineer a mechanical gain that balances the losses 
of the undriven oscillator. This gain-loss balance threshold induces rich sets of collective nonlinear dynamics in the system. 
Our proposal is different from those listed above in at least three points. Indeed, (i) the 
collective dynamics are induced through mechanical gain-loss balance, (ii) only phonons are mediating the energy transport 
between the driven and the undriven resonator, and (iii) only one optomechanical cell is excited and the energy is transferred 
to the rest of the system. The figure of merit of our proposal is to use only one driving laser, and our system can also be used to perform phonon transfer in an array of mechanical resonators. Moreover, by adopting the general case of non-degenerated mechanical resonators, we have identified different sets of synchronized states emerging in our proposal. These synchronized states depend on the mechanical coupling, the frequency mismatch of the two resonators, and the external driving that controls the mechanical gain and the 
optical spring effect. Both in-phase and out-of-phase synchronization 
are present, and the phase transition between these two phenomena happens at a phase flip which has been shown to be related to 
a high enough value of the mechanical gain. Moreover, we have used the quadratic coupling to enhance in-phase synchronization 
process. As these rich collective phenomena can be controlled by only tuning the driving 
field, our proposal appears as an efficient platform to perform applications such as 
phonons transport, phonons processing, and metrology. 
The rest of the work is organized as follows. In Sec. \ref{MoEq}, the model and its dynamical equations are described . 
The emerged collective dynamics are presented in Sec. \ref{Col.dyn}, without the quadratic coupling. The 
frequency mismatch effect together with qualitative explanations of the transitional phases are figured out in 
Section \ref{An.Freq}. Section \ref{Quad} is devoted to the effect of quadratic coupling on the enhancement of synchronization, 
and Sec. \ref{Concl} concludes the work.

\section{Modelling and dynamical equations} \label{MoEq}

\begin{figure*}[tbh]
 \begin{center}
 \resizebox{0.4\textwidth}{!}{
 \includegraphics{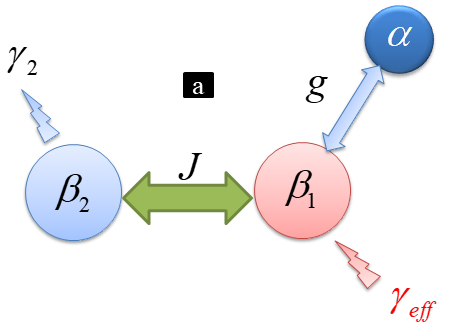}}
 \resizebox{0.4\textwidth}{!}{
 \includegraphics{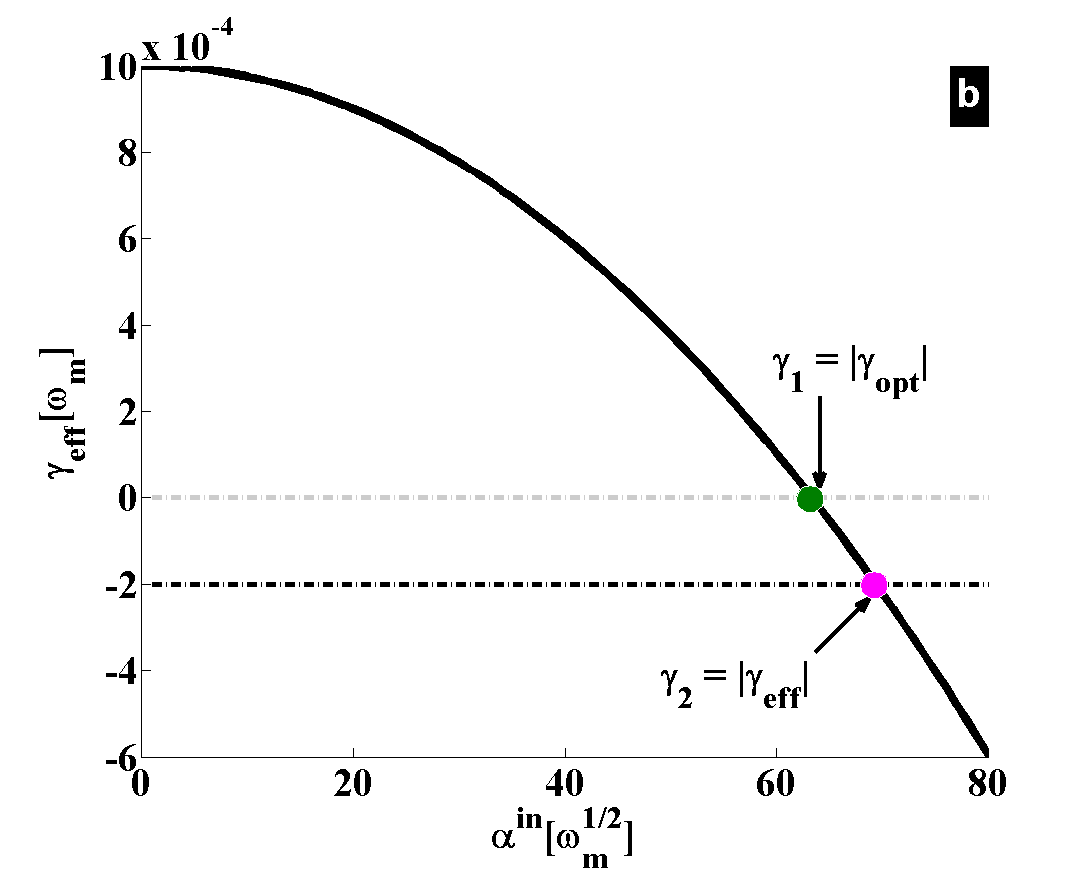}}
 \resizebox{0.4\textwidth}{!}{
 \includegraphics{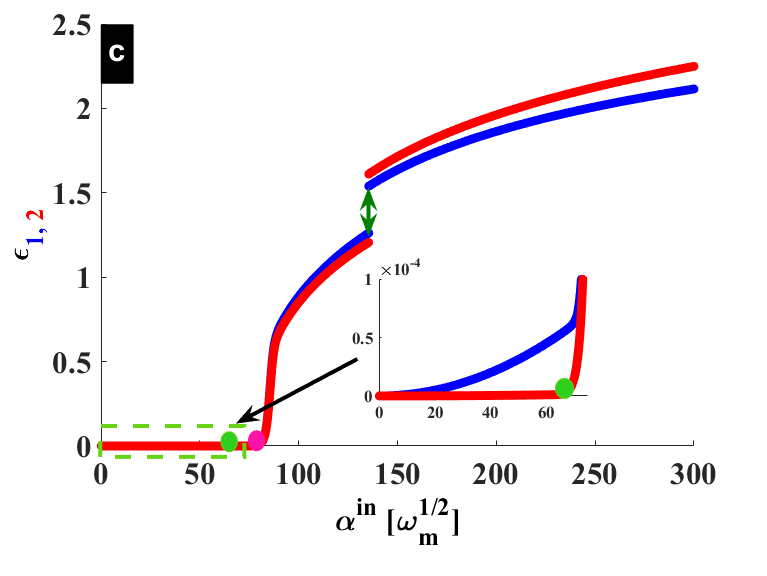}}
 \resizebox{0.4\textwidth}{!}{
 \includegraphics{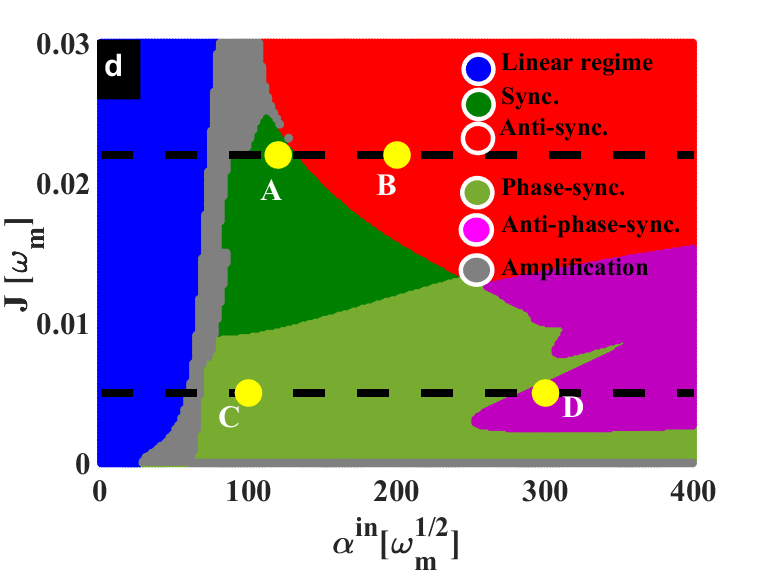}}
 \end{center}
 \caption{(a) Benchmark system consisting of a blue-detuned driven optomechanical cavity, which 
 is mechanically coupled to an undriven mechanical resonator. (b) Effective damping $\gamma_{\rm{eff}}$ versus the driving field $\alpha^{in}$, showing both phonon amplification thresholds for the driven resonator at $\gamma_{1}=\left| \gamma_{\rm{opt}}\right| $ (green dot) and the one for the whole system at $\gamma_{2}=\left| \gamma_{\rm{eff}}\right| $ (pink dot). (c) Normalized mechanical amplitudes of the driven (blue) and the undriven (red) mechanical resonator versus the driving field. The green dashed box is zoomed out in the inset to show phonon lasing of the driven resonator. The vertical double arrow is reminiscent of a phase-flip transition that will be explained in sections \ref{Col.dyn} and \ref{An.Freq}. (d) Diagram displaying collective dynamics in the ($\alpha^{in}$, $J$) parameter's space. Once the losses are balanced by the gain,
 both mechanical resonators carry out different types of dynamics depending on the 
 values of $\alpha^{in}$ and $J$. The coupling has been fixed at $J=2.2\times10^{-2}\omega_m$ for (b) and (c). The other used parameters are,  $\omega_{1}=\omega_m$, $\omega_{2}=1.002\omega_m$, 
 $\gamma_{1}=10^{-3}\omega_m$, $\gamma_{2}=\frac{\gamma_{1}}{5}$, $\kappa=10^{-1}\omega_m$, 
 $g=2.5\times10^{-4}\omega_m$, $\Delta=\omega_m$.}
 \label{fig:Fig1}
 \end{figure*}

Our benchmark system is the one sketched in Fig. \ref{fig:Fig1}a, which 
consists of an optomechanical cavity whose vibrating element is mechanically coupled 
to an undriven auxiliary mechanical resonator. In the rotating frame of the driving fields ($\omega_{p}$), 
the Hamiltonian ($\hbar=1$) describing this system is,

\begin{equation}
H=H_{\rm{OM}}+H_{\rm{int}}+H_{\rm{drive}}, \label{eq1}
\end{equation}
with 
\begin{equation}
\left\{
\begin{array}
[c]{c}
H_{\rm{OM}}=-\Delta a^{\dag}a-ga^{\dag}a(b_{1}^{\dag}+b_{1})+\sum_{j=1,2}\omega_{j}b_{j}^{\dag}b_{j}, \\
H_{\rm{int}}=-J(b_{1}b_{2}^{\dag}+b_{1}^{\dag}b_{2}),  \\
H_{\rm{drive}}=E(a^{\dag}+a).
\end{array}
\right. \label{eq2}
\end{equation}
In this Hamiltonian, $a$ and $b_{j}$ are the annihilation  bosonic field operators describing the optical and mechanical resonators, 
respectively. The mechanical displacements $x_{j}$ are connected to $b_{j}$ as $x_{j}=x_{_{\rm{ZPF}}}(b_{j} +b_{j}^{\dag})$, 
where $x_{_{\rm{ZPF}}}$ is the zero-point fluctuation amplitude of the mechanical resonator. The mechanical frequency 
of the $j^{th}$ resonator is $\omega_{j}$ and  $\Delta=\omega_{p}-\omega_{\rm{cav}}$ is the detuning between the  optical drive ($\omega_{p}$) and the  cavity ($\omega_{\rm{cav}}$) frequencies. 
The mechanical coupling strength between the two mechanical resonators is $J$, and  the optomechanical coupling is $g$. 
The amplitude of the driving pump is $E$. For large photon number in the system, the quantum operators in Eq. (\ref{eq2}) can 
be treated as complex scalar fields, which are defined as the mean 
values of the operators: $\langle a \rangle=\alpha$ and $\langle b_{j} \rangle=\beta_{j}$.
This leads to the following set of classical nonlinear equations of motion for our system,
\begin{equation}
\left\{
\begin{array}{c}
\dot{\alpha}=[i(\Delta+g(\beta_{1}^{\ast}+\beta_{1}))-\frac{\kappa}{2}] \alpha-i\sqrt{\kappa}\alpha^{in}, \\
\dot{\beta}_{1}=-(i\omega_{1}+\frac{\gamma_{1}}{2}) \beta_{1}+iJ\beta_{2}+ig\alpha^{\ast}\alpha, \\
\dot{\beta}_{2}=-(i\omega_{2}+\frac{\gamma_{2}}{2}) \beta_{2}+iJ\beta_{1},
\end{array}
\right.  \label{eq3}
\end{equation}
where optical ($\kappa$) and mechanical ($\gamma_{j}$) dissipations have been added, and the amplitude of the driving 
pump has been substituted as $E=\sqrt{\kappa}\alpha^{\rm{in}}$ in order to account for losses. In this form, the input laser power 
$\rm{P_{in}}$ acts through $\alpha^{in}=\sqrt{\frac{\rm{P_{in}}}{\hbar\omega_{p}}}$. Throughout the work, 
we assume the hierarchy of parameters $\gamma_{j},g\ll \kappa \ll \omega_{j}$, 
similar to those encountered in resolved sideband experiment \cite{Hong.2017}.

Starting from two non-degenerated mechanical resonators, $\omega_{1}\neq \omega_{2}$ and 
$\gamma_{1}\neq \gamma_{2}$, we aim to synchronize their motions. Our strategy is based on 
 engineering and control of the mechanical gain, by driving the optomechanical cavity with a 
blue detuned electromagnetic field. This strategy is sketched on Fig. \ref{fig:Fig1}b where the effective damping ($\gamma_{\rm{eff}}=\gamma_{1}+\gamma_{\rm{opt}}$) is plotted versus the driving field. As we increase the driving strength, this generates optical damping or gain ($\gamma_{\rm{opt}}$) that balances the intrinsic damping ($\gamma_{1}$) of the driven 
mechanical resonator (see green dot in Fig. \ref{fig:Fig1}b) which starts emitting phonons that are mechanically transferred to the second mechanical resonator. This phonon transfer process is revealed in Fig. \ref{fig:Fig1}c (see the inset), which depicts the normalized mechanical amplitudes of both resonators (see Appendix \ref{App.A}) versus the driving field. The blue (red) curve is the amplitude of the driven (undriven) mechanical resonator. It can be seen that the driven resonator (blue curve) is transferring phonons to the undriven one (red), and this transfer process is more pronounced when the optical damping balances the intrinsic damping (compare green dots in Fig. \ref{fig:Fig1}b and Fig. \ref{fig:Fig1}c). 
By further increasing the strength of the driving field, the engineered gain 
increases too and balances the losses of the second mechanical resonator (see magenta dot in Fig. \ref{fig:Fig1}b). 
From this point, the whole system reaches the phonon lasing threshold and both mechanical resonators simultaneously start  emitting phonons (see magenta dot in Fig. \ref{fig:Fig1}c). After this amplification phase, the system 
settles into self-sustained mechanical oscillations regime, above which complex nonlinear behaviours such as 
period doubling and chaos could emerge for strong enough driving strength \cite{Djorwe.2018.PRE}. As we are looking for collective dynamics such as synchronization and frequency locking phenomena, we limit ourselves in this 
work to the self-sustained oscillations regime. In order to optimize the input power needed to balance 
gain and losses in the whole system, we require that the second resonator has a higher quality factor 
than the first one ($\gamma_{1} > \gamma_{2}$). This requirement ensures that our system can reach low-power phonon 
lasing, which is a useful prerequisite for experimental test of our proposal. Indeed, the sooner $\gamma_{1}=\left| \gamma_{\rm{opt}} \right|$ the faster the overall losses is balanced in the whole system by supplying less driven power (compare the green and pink dots in Fig. \ref{fig:Fig1}b and Fig. \ref{fig:Fig1}c).  

\section{Collective dynamical states} \label{Col.dyn}

Beyond the phonon lasing threshold, both mechanical resonators exhibit limit cycle oscillations,  which are 
correlated through their phases and/or their vibrational amplitudes. These correlations lead to different sets of collective nonlinear behaviours, ranging from in-phase to out-of-phase synchronizations with equal or mismatched amplitudes. We have characterized these behaviours from steady state solutions of the mechanical resonators, where all the transient behaviour has died out. The overall dynamics is shown  in Fig. \ref{fig:Fig1}d. The blue area is the linear regime, which is characterized by fixed points  and where there is an energy flow from the driven resonator ($\beta_{1}$) into the undriven one ($\beta_{2}$) as depicted in the inset of Fig. \ref{fig:Fig1}c.  At the gain and loss balance, the parametric instability threshold is reached, and phonon amplification process happens (gray area) until the mechanical amplitudes settle into the self-sustained oscillations regime due to the intrinsic nonlinearity in the system.  Within self-sustained oscillations regime, and depending on the strength of both mechanical coupling $J$ and driving field $\alpha^{in}$, the resonators display several sets of collective dynamical state as shown by different colors beyond the gray area in Fig. \ref{fig:Fig1}d. 
Two quantities have been defined to characterize these dynamical states, the phase difference between the resonators 
and their mismatched amplitudes that is termed error synchronization later on. The instantaneous phase $\phi_i$ of a given resonator is defined as, $\phi_i=\rm{atan} \left(\frac{\rm{Im}(\beta_i)}{\rm{Re}(\beta_i)}\right)$, 
where $\beta$ is a mechanical state variable. The averaged phase difference ($\Delta \phi$)
between both resonator is,
\begin{equation}
 \Delta \phi= \langle|\phi_{i}-\phi_{j}|\rangle \hspace{1.5em} i,j=1,2 \label{eq4}
\end{equation}
where $\langle.\rangle$ denotes an average over time. Similarly, the steady state amplitude of a given resonator 
is captured by $\epsilon\equiv \rm{rms}(\beta_i)$, where ''$\rm{rms}$'' is the root-mean-square value.  We define the 
synchronization error here as a difference between the amplitudes of the resonators,
\begin{equation}
 \rm{Err}= |\epsilon_{i}-\epsilon_{j}| \hspace{1.5em} i,j=1,2 . \label{eq5}
\end{equation}

\begin{figure}[tbh]
 \begin{center}
 \resizebox{0.5\textwidth}{!}{
 \includegraphics{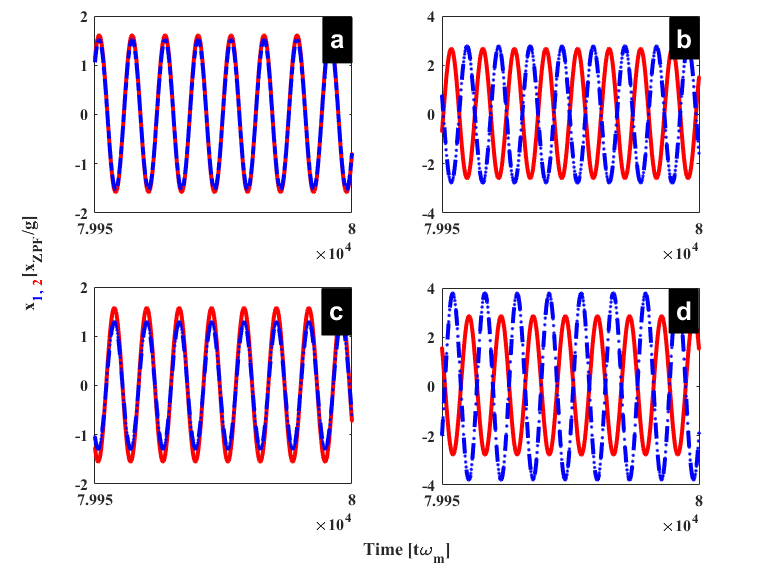}}
\end{center}
 \caption{(a), (b) $\rm{0}$, $\rm{\pi}$-synchronization states. (c), (d) Phase, anti
 phase synchronizations. We highlight that (a), (b), (c) and (d)  are related to the yellow dots A, B, C and D indicated in Fig. \ref{fig:Fig1}d. In (a) and (b), $J=2.2\times10^{-2}\omega_m$ (see upper horizontal dashed line in Fig. \ref{fig:Fig1}d) and $\alpha^{in}= (1.2\times10^2, 2.2\times10^2)\sqrt{\omega_m}$,
  respectively. In (c) and (d),  $J=5\times10^{-3}\omega_m$ (see lower horizontal dashed line in Fig. \ref{fig:Fig1}d) and $\alpha^{in}= (10^2, 3\times10^2)\sqrt{\omega_m}$,
  respectively.}
 \label{fig:Fig2}
 \end{figure}

\begin{figure}[tbh]
 \begin{center}
 \resizebox{0.4\textwidth}{!}{
 \includegraphics{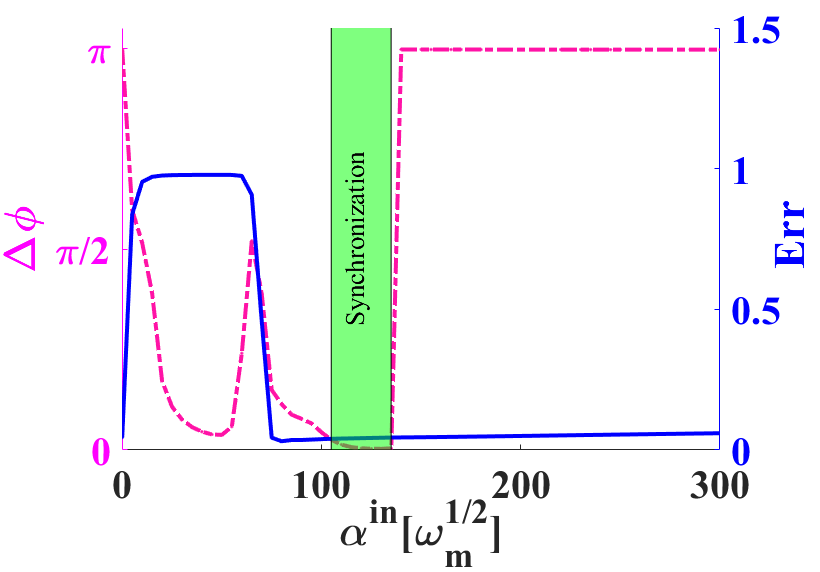}}
\end{center}
 \caption{Phase difference $\Delta \phi$ (dash-dotted curve)  and  error synchronization 
 $\rm{Err}$ (full curve)  between the mechanical resonators for $J=2.2 \times 10^{-2}\omega_m$. The other parameters are the same as in Fig. 1}
 \label{fig:Fig3}
 \end{figure}

Depending on these two quantities, the system exhibits four different dynamical states as depicted beyond the 
gray area in Fig. \ref{fig:Fig1}d. In the green area, one has both $\Delta \phi\sim 0$ and 
$\rm{Err}\sim 0$, resulting in a $\rm{0}$-synchronized state of the mechanical resonators as dynamically depicted 
in Fig. \ref{fig:Fig2}a. It can be clearly seen that both resonators exhibit a similar behaviour in this regime. In the red area however, the resonators are out of phase 
($\Delta \phi\sim \rm{\pi}$), where they oscillate with identical amplitudes ($\rm{Err}\sim 0$) with a phase difference of $\rm{\pi}$. Such a dynamic is depicted in Fig. \ref{fig:Fig2}b, which reveals  $\rm{\pi}$-synchronization of both mechanical resonators over time. In the areas corresponding to a weaker mechanical coupling $J$, the resonators carry out only phase synchronizations. Indeed, phase synchronization is observed in the light-green area where both resonators are in phase ($\Delta \phi\sim 0$), but they have different amplitudes $\rm{Err}\neq 0$. Such a synchronized state is dynamically shown in Fig. \ref{fig:Fig2}c. 
Moreover, the resonators exhibit an anti-phase synchronization in the magenta area, where they evolve out of 
phase from each other with dissimilar vibrational amplitudes ($\rm{Err}\neq 0$) as it can be seen in Fig. \ref{fig:Fig2}d. 
These sets of dynamical states enabled by our proposal reveal its performance in carrying out rich collective 
dynamics compared to those known in the state of the art optomechanical systems. The key point of this performance here is the 
presence of gain and loss, which induces the different phase transitions observed in Fig. \ref{fig:Fig1}d as we will explain later on.
From these displayed collective dynamics, one can deduce two features depending on either the mechanical 
coupling $J$ or the driven strength $\alpha^{in}$ is adjusted. For weak mechanical coupling for instance, the two mechanical resonators 
start exhibiting phase synchronization, which ends up to complete synchronization as $J$ is increasing. 
This is the case for the transition from the light green to the green areas in Fig. \ref{fig:Fig1}d. Similar transition 
happens when the resonators are out of phase as well, where the increase of $J$ adjusts their vibrational amplitudes 
(see switching from magenta to red areas in Fig. \ref{fig:Fig1}d). One can point out also the fact that, 
the driving strength mostly induces switching related to the phase of the resonators. Indeed, as $\alpha^{in}$ 
is increasing, the resonators switch from being in phase ($\Delta \phi\sim 0$) to completely be out of phase ($\Delta \phi\sim \rm{\pi}$). 
This feature can be seen for instance from the transitions between the green and the red areas in Fig. \ref{fig:Fig1}d. This 
switching that is termed later on as phase-flip, is a well-known nonlinear phenomenon characterized by a sudden jump 
of the phase difference roughly from $\rm{0}$ to  $\rm{\pi}$ \cite{Karnatak.2010, Sharma.2012}. We have figured out this phenomenon in Fig. \ref{fig:Fig3}, where the phase difference given in Eq. (\ref{eq4}) (dash-dotted curve) and the synchronization error defined in Eq. (\ref{eq5}) (full curve) are represented for a given $J=2.2 \times 10^{-2}\omega_m$. 
It can be clearly seen that after the amplification regime, both mechanical 
resonators adjust their amplitude and phase, and synchronize for a while within the range 
$(100\lesssim\alpha^{in}\lesssim135)\sqrt{\omega_m}$. Beyond the upper limit of this interval, the phase difference suddenly 
jumps from roughly $\rm{0}$ to $\rm{\pi}$ while the synchronization error remains almost zero ($\rm{Err}\sim 0$). 
This phase-flip phenomenon marks the threshold of the out of phase synchronization, which is induced both 
by gain-loss competition and optical spring effect through an adjustment of the driving strength. We have found that 
the phase-flip transition is likewise revealed through the amplitude jump phenomenon, pointed out by the green double
arrow indicated in Fig. \ref{fig:Fig1}c.  This can be qualitatively explained by an analytical 
approach based on the mechanical eigenmodes of our system. 

\section{Analytical approximations and frequency mismatch effect} \label{An.Freq}

Based on well-known analytical approximations (see Appendix \ref{App.A}), the mechanical eigenmodes of our system 
can be obtained by integrating the optical intracavity field $\alpha (t)$ out of the set of Eq. (\ref{eq3}). 
The resulting eigenmodes yield, 

\begin{equation}
\lambda_{\pm}=  \frac{\omega_{\rm{eff}}+\omega_{2}}{2}-\frac{i}{4}\left(\gamma_{\rm{eff}}+\gamma_{2}
\right) \pm \frac{\sigma}{4},  \label{eq6}
\end{equation}
where  $\omega_{\rm{eff}}=\omega_{1}+\delta \omega_{\rm{opt}}$
and $\gamma_{\rm{eff}}=\gamma_{1}+\gamma_{\rm{opt}}$ are the effective frequency
and  damping of the driven mechanical resonator, respectively. The quantities $\delta \omega_{\rm{opt}}$ and $\gamma_{\rm{opt}}$ 
are respectively the optical spring effect and damping induced by the driving field, and they are given by,
\begin{equation}
\delta \omega_{\rm{opt}}=-\frac{2\kappa (g\alpha^{in})^{2}}{\omega
	_{\rm{lock}}\epsilon_{1}}\rm{Re}\left(\sum_{n}\frac{J_{n+1}\left(
	-\epsilon_{1}\right) J_{n}\left(-\epsilon_{1}\right)}{
	h_{n+1}^{\ast}h_{n}}\right),  \label{eq.7}
\end{equation}

and 

\begin{equation}
\gamma_{\rm{opt}}=\frac{2(g\kappa \alpha ^{\rm{in}})^{2}}{\epsilon_{1}}
\sum_{n}\frac{J_{n+1}\left(-\epsilon_{1}\right) J_{n}\left(
	-\epsilon_{1}\right)}{\left\vert h_{n+1}^{\ast}h_{n}\right\vert^{2}},  \label{eq.8}
\end{equation} 
where $\epsilon_{1}=\frac{2g\rm{Re}(A_{1})}{\omega_{\rm{lock}}}$ is a normalized amplitude, with $A_{1}$  and $\omega_{\rm{lock}}$ the mechanical amplitude of the first resonator and its locked frequency, respectively. 
$J_{n}$ is the Bessel function of the first kind of order $n$,  $h_{n}=i\left(n\omega_{\rm{lock}}-\tilde{\Delta}\right) 
+\frac{\kappa}{2}$, and $\tilde{\Delta}=\Delta_{1}+2g(\bar{\beta}_{1})$ 
the effective detuning. In Eq. (\ref{eq6}), $\sigma$ is a complex quantity defined as,
\begin{equation}
\sigma = \sqrt{16J^{2}+[2(\omega_{eff}-\omega_{2}) 
	+i(\gamma_{eff}-\gamma_{2})]^2}. \label{eq.9}
\end{equation}
whose real and imaginary part read,  
\begin{equation}
\rm{Re}\left(\sigma \right)=\sqrt{\frac{\sqrt{\chi^{2} + (4\Delta\omega_{\rm{eff}}\Delta\gamma_{\rm{eff}})^{2}}+\chi}{2}},  \label{eq7}
\end{equation}
and 
\begin{equation}
\rm{Im}\left(\sigma \right)=\sqrt{\frac{\sqrt{\chi^{2} + (4\Delta\omega_{\rm{eff}}\Delta\gamma_{\rm{eff}})^{2}}-\chi}{2}},  \label{eq8}
\end{equation}
where  $\chi=16J^{2}+4\Delta\omega_{\rm{eff}}^{2}-\Delta\gamma_{\rm{eff}}^{2}$ with 
$\Delta\omega_{\rm{eff}}=\omega_{\rm{eff}}-\omega_{2}$ and $\Delta\gamma_{\rm{eff}}=\gamma_{2}-\gamma_{\rm{eff}}$.
From Eqs.(\ref{eq6})-(\ref{eq8}), it appears that the mechanical supermodes can be characterized 
by their frequencies ($\rm{Re}\left(\lambda_{\pm} \right)=\omega_{\pm}$) and dampings 
($\rm{Im}\left(\lambda_{\pm} \right)=\gamma_{\pm}$) given by,

\begin{equation}
\omega_{\pm}= \frac{\omega_{\rm{eff}}+\omega_{2}}{2}  \pm \frac{\rm{Re}\left(\sigma \right)}{4},  \label{eq9}
\end{equation}
and
\begin{equation}
\gamma_{\pm}= -\frac{\left(\gamma_{\rm{eff}}+\gamma_{2} \right)}{4}  \pm \frac{\rm{Im}\left(\sigma \right)}{4}.  \label{eq10}
\end{equation}

\begin{figure}[tbh]
	\begin{center}
		\resizebox{0.5\textwidth}{!}{
			\includegraphics{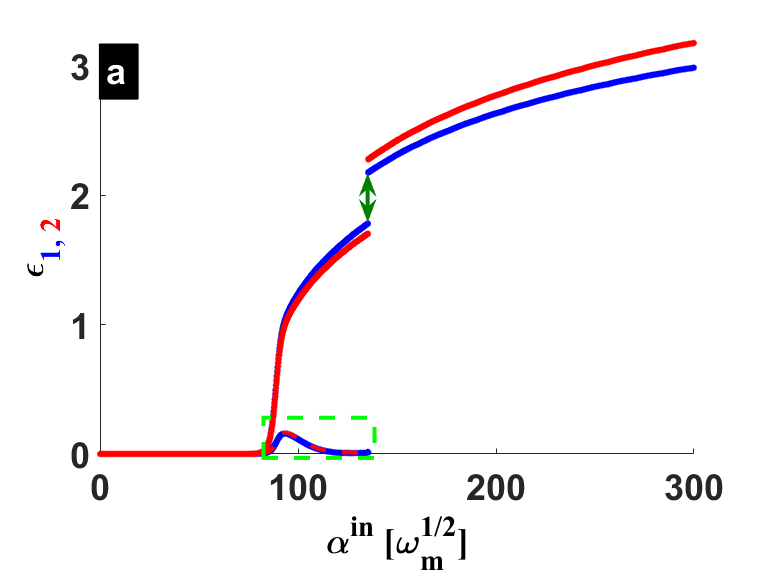}}
		\resizebox{0.5\textwidth}{!}{
			\includegraphics{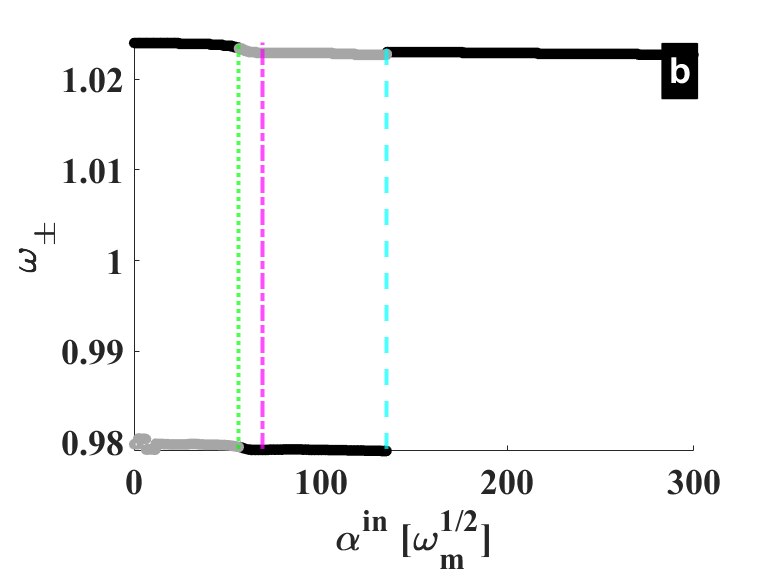}}
	\end{center}
	\caption{Mechanical supermodes obtained by Fast Fourier Transforming the data of Fig. \ref{fig:Fig1}c. (a) Normalized mechanical amplitudes ($\epsilon_{1,2}$) of the resonators. These amplitudes can be viewed as two upper ($\epsilon_{+}$) and two lower supermode amplitudes with the corresponding frequencies $\omega_{\pm}$. (b) Supermode frequencies ($\omega_{\pm}$). The black color corresponds to the supermode oscilating with the highest amplitude, and the gray color is related to the supermode oscillating with the lower amplitude (see further details in Fig. \ref{fig:Fig5}). The vertical lines correspond to: (i) the phonon lasing of the driven resonator (green line), (ii) the phonon lasing in the whole system (magenta line) and (iii) the phase-flip transition (cyan line). These curves are plotted versus $\alpha^{\rm{in}}$ for a fixed $J=2.2\times10^{-2}\omega_m$, and the other parameters are as in Fig. \ref{fig:Fig1}.}
	\label{fig:Fig4}
\end{figure}

In order to reveal the mechanical supermodes involved in our study, we can fast Fourier transform (FFT) the steady state solutions of the mechanical resonators, where all the transient behaviour has died out. From this FFT data, we collect the dominant peaks together with their corresponding frequencies, which represent the amplitudes and the frequencies of the supermodes, respectively. Applying this procedure to the data of Fig. \ref{fig:Fig1}c for instance, has led to the supermodes depicted in  Fig. \ref{fig:Fig4}.

In Fig. \ref{fig:Fig4}a, it can be clearly seen that each mechanical resonator is characterized by two supermodes ($\epsilon_{\pm}$) which are well captured through their supermode frequencies ($\omega_{\pm}$) in Fig. \ref{fig:Fig4}b. More importantly, the lower supermode amplitudes ($\epsilon_{-}$) died out exactly where the phase-flip and amplitude jump phenomena happen (see green rectangle in Fig. \ref{fig:Fig4}a). Furthermore, the supermode frequencies merge to a single frequency ($\omega_{\rm{lock}}$) at the same transitional point (Fig. \ref{fig:Fig4}b), which clearly reveals a frequency locking effect and can be explained by the mode competition phenomenon as well \cite{Sheng.2020}.

From Eqs.(\ref{eq7})-(\ref{eq10}) together with the frequency dynamics figured out in Fig. \ref{fig:Fig4}b, one can analytically predict the transitional phases involved in our system. These transitions are indicated by the vertical lines in Fig. \ref{fig:Fig4}b and they refer to the phonon lasing threshold phenomena (green and magenta lines)  and the phase-flip effect (cyan line). Since the first transition (green line) is not directly related to the synchronization feature, the specific cases of the global phonon lasing (magenta line) and phase-flip are detailed in the following. For degenerated mechanical resonators for instance ($\omega_{1}=\omega_{2}$), one gets $\Delta\omega_{\rm{eff}}\sim0$ for a weak driving field. At the balanced gain and loss, it results that $\rm{Im}\left(\sigma \right)=0$ and $\rm{Re}\left(\sigma \right)=\sqrt{16J^{2}-\Delta\gamma_{\rm{eff}}^{2}}$. This leads to a well-known exceptional point (EP) at $J=\frac{\Delta\gamma_{\rm{eff}}}{4}$, which can be used for mass-sensing \cite{Djorwe.2019.sensor}  and it also marks the threshold of phonon lasing where the mechanical resonators exhibit self-sustained limit cycle oscillations \cite{Djorwe.2018.PRE}. 

   \begin{figure}[tbh]
		\begin{center}
			\resizebox{0.5\textwidth}{!}{
				\includegraphics{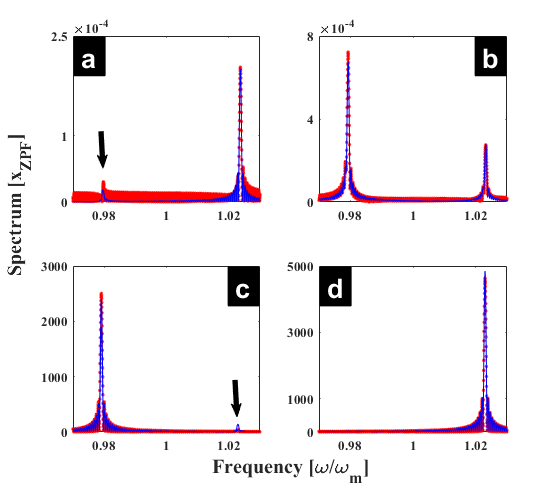}}
		\end{center}
		\caption{Different spectra corresponding to the states within the areas delimited by vertical lines in Fig. \ref{fig:Fig4}. (a) For $\alpha ^{\rm{in}}=54 \sqrt{\omega_m}$ and is before the green vertical line. (b) For $\alpha ^{\rm{in}}=65\sqrt{\omega_m}$ and is between the green and magenta vertical lines. (c) For $\alpha ^{\rm{in}}=90\sqrt{\omega_m}$ and is between the magenta and cyan vertical lines. (d) For $\alpha ^{\rm{in}}=150\sqrt{\omega_m}$ and is beyond the cyan vertical line. Despite the dynamical switching of the peaks between the frequencies $\omega_{\pm}$, attention should be also paid on the growing intensity from (a) to (d), which reveals the action of $\alpha ^{\rm{in}}$.}
		\label{fig:Fig5}
	\end{figure}
In our proposal however, there is a frequency mismatch $\delta \omega = \omega_{2}-\omega_{1}$, which acts as a perturbation on the EP. At the balanced gain-loss, the requirement of EP ($\rm{Re}\left(\sigma \right)=\rm{Im}\left(\sigma \right)=0$) is fulfilled in our system providing that $\chi=0$ and $\Delta\omega_{\rm{eff}}\Delta\gamma_{\rm{eff}}=0$. The former condition is satisfied for 
$4J^{2}+\Delta\omega_{\rm{eff}}^{2}=\gamma_{\rm{2}}^{2}$, while the latter condition can not be satisfied at this weak driving field regime, since $\Delta\gamma_{\rm{eff}}\simeq2\gamma_{\rm{2}}$ and $\Delta\omega_{\rm{eff}}=\delta\omega_{\rm{opt}}-\delta\omega$.  Therefore, one obtains
\begin{align}
\omega_{\pm}&= \frac{\omega_{\rm{eff}}+\omega_{2}}{2}  \pm \sqrt{\frac{\Delta\omega_{\rm{eff}}\Delta\gamma_{\rm{eff}}}{8}} \nonumber\\
  &\simeq\frac{1}{2}\left(\omega_{\rm{eff}}+\omega_{2}\pm\sqrt{\gamma_{2}\Delta\omega_{\rm{eff}}}\right), \label{eq11}
\end{align}
and
\begin{align}
\gamma_{\pm}&=-\frac{\left(\gamma_{\rm{eff}}+\gamma_{2} \right)}{4}  \pm \sqrt{\frac{\Delta\omega_{\rm{eff}}\Delta\gamma_{\rm{eff}}}{8}} \nonumber\\
&\simeq\frac{1}{2}\left(-\frac{\left(\gamma_{\rm{eff}}+\gamma_{2} \right)}{2}\pm \sqrt{\gamma_{2}\Delta\omega_{\rm{eff}}} \right)\label{eq12}
\end{align}
which clearly reveals the effect of the frequency mismatch ($\delta\omega$) on getting $\rm{Re}\left(\sigma \right)=\rm{Im}\left(\sigma \right)=0$. From Eq.(\ref{eq11}) and Eq.(\ref{eq12}), it results that two supermodes are still possible although the losses are balanced by the gain in the whole system as shown in Fig. \ref{fig:Fig4}. In order to reach the transition point where the phase-flip happens, one  needs to further increase the driving strength $\alpha ^{\rm{in}}$ (see Fig. \ref{fig:Fig4}). However, increasing $\alpha ^{\rm{in}}$ leads to 
the increase of both $\delta\omega_{\rm{opt}}$ and $\gamma_{\rm{opt}}$ as it appears in Eq.(\ref{eq.7}) and Eq.(\ref{eq.8}). For enough driving strength, one reaches $\delta\omega_{\rm{opt}}=\delta\omega$ leading to $\Delta\omega_{\rm{eff}}=0$. Therefore, the previous unfulfilled condition ($\Delta\omega_{\rm{eff}}\Delta\gamma_{\rm{eff}}=0$) is now satisfied and both Eq.(\ref{eq9}) and Eq.(\ref{eq10}) yield, 

  \begin{equation}
  \omega_{\pm}= \frac{\omega_{\rm{eff}}+\omega_{2}}{2},  \label{eq13}
  \end{equation}
  and
  \begin{equation}
  \gamma_{\pm}= -\frac{\left(\gamma_{\rm{eff}}+\gamma_{2} \right)}{4}  \pm \sqrt{\chi_{0}},  \label{eq14}
  \end{equation}
since $J<\frac{\Delta\gamma_{\rm{eff}}}{4}$ at this strong driving regime and we have defined $\chi_{0}=\Delta\gamma_{\rm{eff}}^{2}-16J^{2}$. 
These equations reveal that one supermode has vanished ($\rm{Re}\left(\sigma \right)=0$) and this means that each mechanical resonator is now oscillating with a single frequency given by Eq.(\ref{eq13}). From this phase-flip transition, each resonator carries out oscillations that decay according to one of the dissipations given in Eq.(\ref{eq14}).

From the above discussion, it follows that both mechanical gain and optical spring are the key points for the 
collective phenomena arising in our proposal. Owing to the fact that these quantities are tuned through 
the driving field, our proposal appears as a nice platform to generate collective phenomena, where the system 
is wholly controlled externally. To bear in mind the dynamics of the supermodes, we have provided the spectra given in Fig. \ref{fig:Fig5}. These spectra capture the energy transfer between the supermodes for the different situations presented in Fig. \ref{fig:Fig4}b. The spectrum in Fig. \ref{fig:Fig5}a depicts the dynamics before the green line, Fig. \ref{fig:Fig5}b shows the situation between the green and magenta lines, the case within the area bounded by the magenta and cyan line is given in Fig. \ref{fig:Fig5}c, while the case beyond the cyan line is shown by the spectrum in Fig. \ref{fig:Fig5}d. It can be seen that the energy is first transferred from the higher frequency supermode ($\omega_{+}$) to the lower one ($\omega_{-}$), and from the phase-flip transition, the whole energy is swapped back to the supermode $\omega_{+}$. This feature can be further exploited to perform phonon information processing between the mechanical resonators.
\begin{figure}[tbh]
 \begin{center}
 \resizebox{0.5\textwidth}{!}{
 \includegraphics{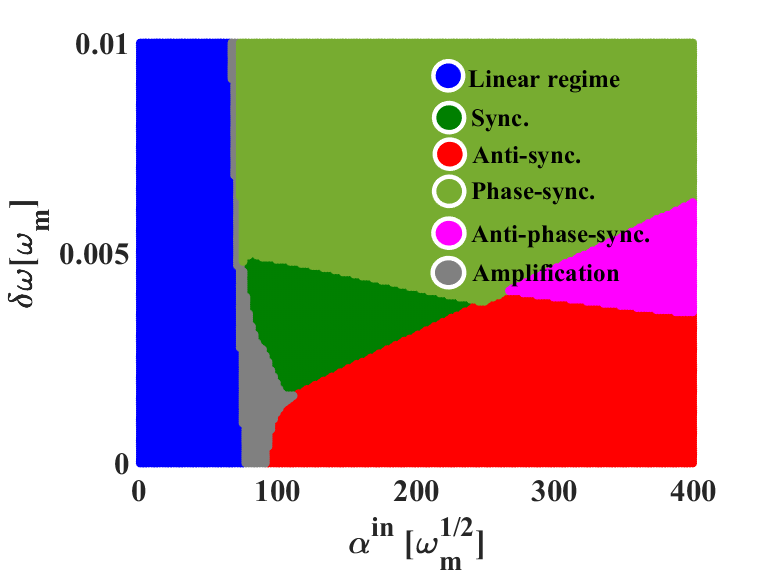}}
\end{center}
 \caption{Numerical diagram displaying collective dynamics in the 
 ($\alpha^{in}$, $\delta \omega$) parameter's space for $J=2.2\times10^{-2}\omega_m$. 
 The other parameters are as in Fig. \ref{fig:Fig1}.}
 \label{fig:Fig6}
 \end{figure}

Based on the fact that the frequency mismatch has been needed to explain the frequency locking effect through $\rm{Re}\left(\sigma \right)=0$ or $\delta\omega_{\rm{opt}}=\delta\omega$, a particular attention must be paid on it. Therefore, the effect of  $\delta \omega$ on the organization of the collective dynamics is shown in Fig. \ref{fig:Fig6}, for a fixed coupling strength
$J=2.2\times10^{-2}\omega_m$. It can be seen that, the frequency mismatch very weakly shifts the lasing threshold (transition between blue and gray areas), compared to the coupling strength $J$ (as in Fig. \ref{fig:Fig1}d). This feature simply reveals the robustness of the amplification process at the vicinity of the EP, against moderate frequency mismatches as aforementioned. From Fig. \ref{fig:Fig6}, 
this robustness is roughly preserved up to $1\%$ of frequency mismatch.   
However, just above the lasing threshold, the mismatch 
induces new reorganization of collective dynamics. For a weak mismatch ($\delta \omega<2\times10^{-3}\omega_m$), 
$\pi$-synchronization is the dominant dynamics. This is expected and is explained by the fact that, weak driving strength 
is required for the mechanical gain and optical spring effect to reverse the sign of $\chi$ and to compensate the mismatch, 
respectively. For large mismatch however, strong driving strength is required 
to generate large gain and optical spring effect. Therefore, it is hard to reach $\pi$-synchronization in such a configuration as 
it can be seen for $\delta \omega\gtrsim 5\times10^{-3}\omega_m$, where only phase synchronization emerges.
For intermediate values of the mismatch ($2\times10^{-3}\omega_m\lesssim\delta \omega\lesssim4\times10^{-3}\omega_m$), both 
resonators can synchronize for a while before switching to out of phase synchronization. Owing to the difficulties 
of engineering identical mechanical resonators in micro/nano-fabrication technologies, 
it could be interesting to seek an enhancement strategy of the observed collective phenomena as presented in the next section.

\section{Quadratic coupling enhances in-phase dynamics} \label{Quad}
 
\begin{figure}[tbh]
 \begin{center}
 \resizebox{0.4\textwidth}{!}{
 \includegraphics{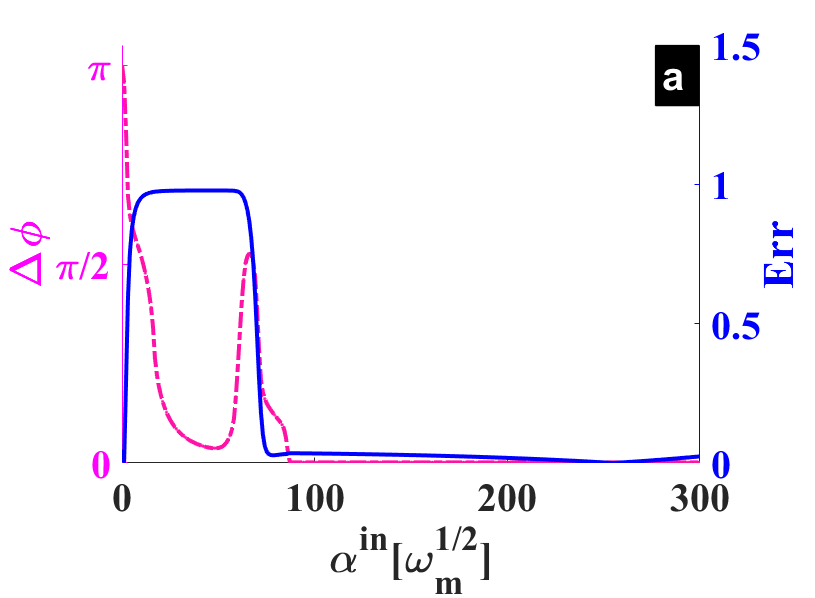}}
\resizebox{0.4\textwidth}{!}{
	\includegraphics{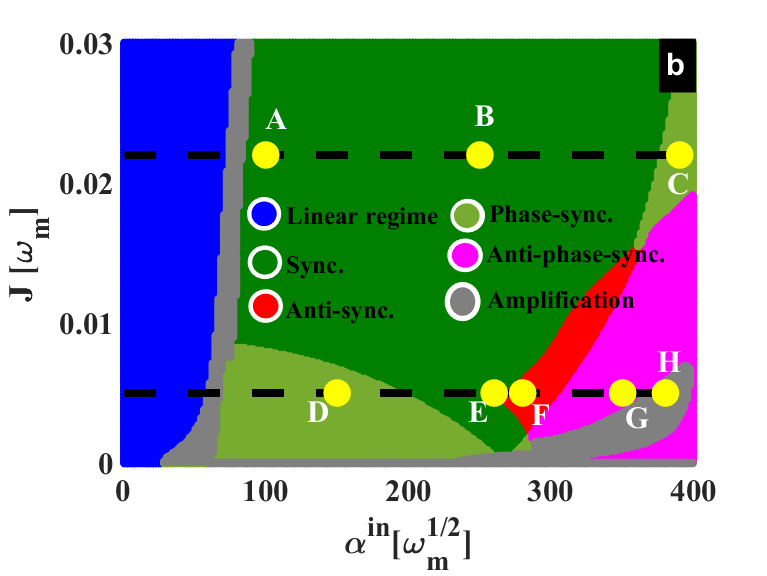}}
\end{center}
 \caption{Enhancement effect of the quadratic coupling on in-phase synchronization.  
 (a) Phase difference $\Delta \phi$ (dash-dotted curve) and error synchronization 
 $\rm{Err}$ (full curve) between the mechanical resonators for $J=2.2\times10^{-2}\omega_m$. 
 (b) Numerical diagram displaying collective dynamics
 in the ($\alpha^{in}$, $J$) parameter's space. Quadratic coupling of $g_{ck}/g=-10^{-3}$ has 
 been accounted,  and the other parameters are the same as in Fig. \ref{fig:Fig1}.}
 \label{fig:Fig7}
 \end{figure}

In order to assist the role played by the optical spring effect and to enhance the effect of the mechanical gain, 
we have introduced the quadratic coupling ($g_{\rm{ck}}$) in the system. This well-known nonlinearity can be generated by 
inserting the driven mechanical resonator inside the optical cavity for instance \cite{Paraiso.2015}. The resulted system can be 
thought as a membrane-in-the-middle setup where the moving element is mechanically coupled to the undriven mechanical 
resonator. In such a system, the frequency of the driven mechanical system yields $\tilde{\omega}_{1} =\omega_{1}-g_{ck}\alpha^{\ast}\alpha$, 
which can be tuned through the driving field (see Appendix \ref{App.B}). This frequency control is useful to bring closer the frequencies of the non-degenerated mechanical resonators involved in our system, even at the vicinity of an EP 
\cite{Djorwe.2019.sensor}.  This can be seen by comparing Fig. \ref{fig:Fig7}a to Fig. \ref{fig:Fig3}.  Indeed, $\rm{0}$-synchronization is established before $\alpha^{in}= 10^2\sqrt{\omega_m}$ in Fig. \ref{fig:Fig7}a (see dash-dotted curve) where the quadratic term is $g_{ck}/g=-10^{-3}$, whereas this happens above such a value in Fig. \ref{fig:Fig3} where $g_{ck}/g=0$. It can be also seen that $\rm{0}$-synchronization is wider in  Fig. \ref{fig:Fig7}a than in Fig. \ref{fig:Fig3}, revealing the effect of the quadratic coupling in enhancing in-phase synchronization. This enhancement effect of the quadratic term is depicted 
in Fig. \ref{fig:Fig7}b, where in-phase synchronization (green area) is clearly improved. Moreover, by comparing 
the diagrams depicted in Fig. \ref{fig:Fig7}b and Fig. \ref{fig:Fig1}d, it results that quadratic coupling 
also reorganizes localization of the different dynamical states. We have checked the dynamics for the labelled points in 
Fig. \ref{fig:Fig7}b, and the corresponding dynamical states satisfy the expected behaviours (see Fig. \ref{fig:Fig8}). 
Furthermore, the dynamical state of the gray area intertwined in the magenta zone has been figured out, and its corresponds 
to a strong coupling regime where both resonators exchange energy through Rabi oscillations (Appendix \ref{App.B}).  
 
\section{Conclusion} \label{Concl} 
We have investigated collective dynamics in a blue-detuned optomechanical cavity that is mechanically 
coupled to an undriven mechanical resonator. When the optomechanically engineered mechanical gain  
balances the losses of the undriven resonator, phonon lasing threshold happens and both resonators 
simultaneously exhibit self-sustained limit cycles, leading to interesting sets of collective dynamics. 
Depending on the mechanical coupling and the external driving field, we have observed in-phase and out-of-phase 
synchronizations that mainly result from process of driving that induces both mechanical gain and optical spring. 
Qualitative explanations of the main phase transitions arising in our proposal have been provided, based on mechanical 
eigenvalues evaluated through analytical approximations. Furthermore, we have used quadratic  optomechanical coupling 
to enhance in-phase synchronization between the non-degenerated mechanical resonators involved in the system. 
Our work opens new avenues towards collective dynamics in an array of mechanically coupled resonators by only driving 
one of them. This scheme can be extended to related systems including, electromechanical and superconducting microvawe setups.

\section*{Acknowledgments}

This work was supported by the European Commission FET OPEN H2020
project PHENOMEN-Grant Agreement No. 713450 and the funding from the European Union’s Horizon 2020 research and innovation programme under the Marie Skłodowska-Curie grant agreement No. 754510, under which part of this work was completed.

\appendix \label{App}

\section{Mechanical supermodes} \label{App.A}  
Our aim here is to give few details about the steps leading to the mechanical supermodes mentioned in section \ref{An.Freq}. To avoid reproducing some expressions that are already explained above, we start from Eq.\ref{eq3} in the main text which is the classical nonlinear set of equations describing our proposal.
Only one mechanical resonator ($\beta_{1}$) is driven by a blue-detuned electromagnetic field while the second ($\beta_{2}$) is just mechanically coupled  to it.  Therefore, the following optomechanical transformation will be only applied to $\beta_{1}$. However, we underline that both resonators can oscillate for enough driving strength, owing to the mechanical coupling $J$ as explained in the main text.  Once the limit cycles regime is reached, this driven resonator carries out oscillations whose amplitudes change
only slowly over time. Thus, we  solve the equation 
for $\alpha$ assuming a fixed amplitude for the mechanical oscillations, and then use the
result to solve the equation for $\beta_{1}$. Under this assumption, the mechanical oscillation can be described by the ansatz \cite{Djorwe.2018.PRE},

\begin{equation}
\beta_{j}(t)=\bar{\beta_{j}}+A_{j}\exp(-i\omega_{\rm{lock}}t),
\label{S1}
\end{equation}
where $\bar{\beta_{j}}$ is a constant shift in the origin of the resonator
and the amplitude $A_{j}$ is taken to be a slowly varying function of time. 
In such regime of single frequency ansatz (see Fig. \ref{fig:Fig4}b for instance), we have denoted the locked frequency by $\omega_{\rm{lock}}$.
We substitute this ansatz into the equation for $\alpha$, and use the
assumption of a slowly evolving amplitude to solve it, first neglecting the
time dependence of $A_{j}$. We then obtain the intracavity field
in the form,

\begin{equation}
\alpha(t)=e^{-i\theta\left(t\right)}\sum_{n}\alpha
_{n}e^{in\omega_{\rm{lock}}t}.  \label{S2}
\end{equation}
The phase is $\theta\left(t\right) =-\epsilon_{1}\sin \omega_{\rm{lock}}t$
and the amplitudes of the different harmonics of the optical field are,
\begin{equation}
\alpha_{n}=-i\sqrt{\kappa}\alpha^{in}\frac{J_{n}\left(
	-\epsilon_{1}\right)}{h_{n}},  \label{S3}
\end{equation}
where $\epsilon_{1}=\frac{2g\rm{Re}(A_{1})}{\omega_{\rm{lock}}}$ , $\tilde{
	\Delta}=\Delta+2g\rm{Re}(\bar{\beta_{1}})$, $h_{n}=i\left(
n\omega_{\rm{lock}}-\tilde{\Delta}\right) +\frac{\kappa }{2}$ and $J_{n}$ is
the Bessel function of the first kind of order $n$.

As we are interested in the regime of limit cycles of the resonators, a
rotating wave approximation can be made in which we drop all the terms (in
the mechanical dynamics) except the constant one and the term oscillating at
$\omega_{\rm{lock}}$. Hence, we substitute Eq. (\ref{S2}) in the equation for $
\beta_{1}$ (see Eq. (\ref{eq3})) which, by equating constant terms, leads
to the zero-frequency components,

\begin{equation}
\bar{\beta}_{1}=\frac{1}{\omega_{1}-i\frac{\gamma_{m}}{2}}\left(g\kappa
\sum_{n}\frac{\left(\alpha^{in}J_{n}\left(-\epsilon\right)
	\right)^{2}}{\left\vert h_{n}\right\vert ^{2}}+J\bar{\beta}_{2}\right), \label{S4}
\end{equation}
that induce a shifts of the cavity frequencies,
\begin{equation}
\delta=2g\rm{Re}(\bar{\beta}_{1}).  \label{S5}
\end{equation}

The equations of motion for the oscillating part of $\beta _{j}$ are deduced
from $\beta_{r}^{j}(t)=\beta_{j}(t)-\bar{\beta}_{j}\equiv A_{j}\exp
(-i\omega_{\rm{lock}}t)$ and read,
\begin{equation}
\left\{
\begin{array}{c}
\dot{\beta}_{r}^{1}(t)=-i\left(\omega_{1}+\delta \omega_{\rm{opt}}\right) \beta
_{r}^{1}-\frac{\gamma_{1}+\gamma_{opt}}{2}\beta_{r}^{1}+iJ\beta
_{r}^{2} \\
\dot{\beta}_{r}^{2}(t)=-i\omega_{2}\beta_{r}^{2}-\frac{\gamma_{2}}{2}\beta_{r}^{2}+iJ\beta_{r}^{1}
\end{array}
\right.  \label{S6}
\end{equation}

Here the optical spring effect $\delta \omega_{\rm{opt}}$ and the optical damping $\gamma_{\rm{opt}}$ coming both from the average dynamics of the cavity are
given by Eq.\ref{eq.7} and Eq.\ref{eq.8} in the main text, respectively.  
From Eq.(\ref{S6}), it is possible to define an effective Hamiltonian in order
to figure out supermodes involved in the system. Such supermodes will be
deduced from the eigenmodes of the effective model, describing the mechanical resonators. 
Indeed, the real parts of the eigenmodes give the eigenfrequencies of the coupled system while their
imaginary parts stand for the dissipations rate of the system. In the limit cycles
regime, the constant shift $\bar{\beta}_{j}$ is weak compared to the
amplitude of the mechanical resonator ($\bar{\beta}_{j}\ll A_{j}$).
This means that $\beta_{j}(t)\cong \beta_{r}^{j}(t)$, and Eq.(\ref{S6})
can be assumed as a set of equations describing the effective system that reads,

\begin{equation}
\left\{
\begin{array}{c}
\dot{\beta}_{1}=-\left(i\omega_{\rm{eff}}+\frac{\gamma_{\rm{eff}}}{2}
\right) \beta_{1}+iJ\beta_{2}, \\
\dot{\beta}_{2}=-\left(i\omega_{2}+\frac{\gamma_{2}}{2}\right) \beta_{2}+iJ\beta_{1},
\end{array}
\right.  \label{S10}
\end{equation}
where $\omega_{\rm{eff}}=\omega_{1}+\delta \omega_{\rm{opt}}$ and $\gamma_{\rm{eff}}=\gamma_{1}+\gamma_{\rm{opt}}$ define the effective
frequencies and the effective damping, respectively.

Furthermore, Eq.(\ref{S10}) can be rewritten in the compact form,

\begin{equation}
\partial t\Psi =-iH_{\rm{eff}}\Psi
\end{equation}
with the effective Hamiltonian,
\begin{equation}
H_{\rm{eff}}=
\begin{bmatrix}
\omega_{\rm{eff}}-i\frac{\gamma_{\rm{eff}}}{2} & -J \\
-J & \omega_{2}-i\frac{\gamma_{2}}{2}
\end{bmatrix}
\label{S11}
\end{equation}
and the state vector $\Psi =\left(\beta_{1},\beta_{2}\right)^{T}$.

The eigenvalues of the Hamiltonian given in Eq.(\ref{S11}) are obtained by
solving the equation,

\begin{equation}
\det \left( H_{\rm{eff}}-\lambda I\right) =0,
\end{equation}
and that yields to the following eigenvalues $\lambda_{-}$ and $\lambda_{+}$,
\begin{equation}
\lambda_{\pm}\simeq  \frac{\omega_{\rm{eff}}+\omega_{2}}{2}-\frac{i}{4}\left(\gamma _{\rm{eff}}+\gamma_{2}\right) \pm \frac{\sigma}{2}.  \label{S12}
\end{equation}
as given by Eq. (\ref{eq6}) in the main text with the complex quantity $\sigma$ defined by Eq. (\ref{eq.9}) as well. 

 The frequencies and the dissipations of the supermodes discussed in the main text are given by the real and imaginary parts of $\lambda_{\pm}$, respectively

\begin{equation*}
\omega_{\pm}=\rm{Re} \left(\lambda_{\pm}\right) \text{ and }\gamma
_{\pm }=\rm{Im} \left(\lambda_{\pm}\right).
\end{equation*}

\section{Quadratic coupling and dynamical states} \label{App.B}
By taking into account the quadratic coupling (or cross-Kerr) in the model, the Hamiltonian becomes,
\begin{equation}
H=H_{\rm{OM,ck}}+H_{\rm{int}}+H_{\rm{drive}}, \label{eqS1}
\end{equation}
with 
\begin{equation}
\left\{
\begin{array}
[c]{c}
H_{\rm{OM,ck}}=-\Delta a^{\dag}a+\sum_{j=1,2}\omega_{j}b_{j}^{\dag}b_{j}\\ 
-ga^{\dag}a(b_{1}^{\dag}+b_{1})-g_{ck}a^{\dag}ab^{\dag}b ,\\
H_{\rm{int}}=-J(b_{1}b_{2}^{\dag}+b_{1}^{\dag}b_{2}),  \\
H_{\rm{drive}}=E(a^{\dag}+a),
\end{array}
\right. \label{eqS2}
\end{equation}
where $\rm{ck}$ stands for ''cross-Kerr'' and accounts for the quadratic coupling. 
This leads to the following classical set of nonlinear equations,  
\begin{equation}
\left\{
\begin{array}{c}
\dot{\alpha}=[i(\Delta+g(\beta_{1}^{\ast}+\beta_{1})+g_{ck}\beta_{1}^{\ast}\beta_{1})-\frac{\kappa}{2}] \alpha-i\sqrt{\kappa}\alpha^{in}, \\
\dot{\beta}_{1}=-(i\tilde{\omega}_{1}+\frac{\gamma_{1}}{2}) \beta_{1}+iJ\beta_{2}+ig\alpha^{\ast}\alpha, \\
\dot{\beta}_{2}=-(i\omega_{2}+\frac{\gamma_{2}}{2}) \beta_{2}+iJ\beta_{1},
\end{array}
\right.  \label{eqS3}
\end{equation}
where $\tilde{\omega}_{1} =\omega_{1}-g_{\rm{ck}}\alpha^{\ast}\alpha$ is the optically tunable mechanical frequency mentioned 
in the main text.

\begin{figure*}[tbh]
 \begin{center}
 \resizebox{1.0\textwidth}{!}{
 \includegraphics{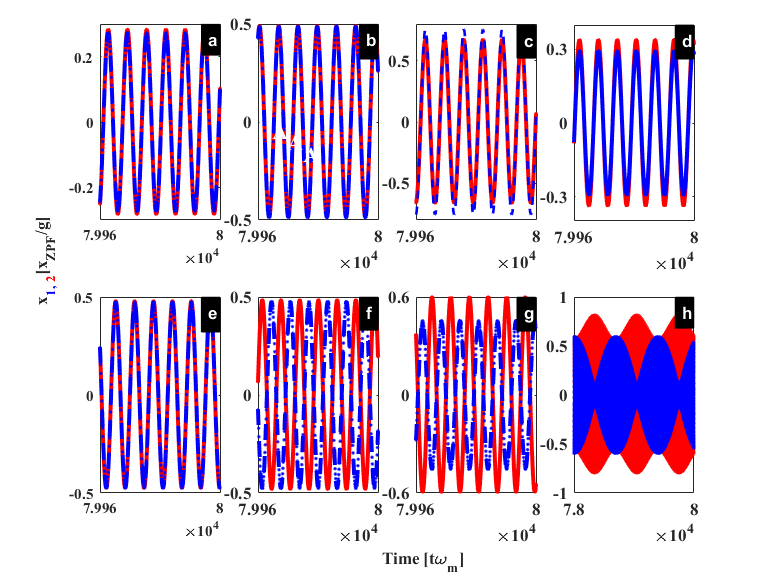}}
\end{center}
 \caption{Dynamical states. (a) to (h) correspond to the points labelled A to H in  Fig. \ref{fig:Fig7}b.}
 \label{fig:Fig8}
 \end{figure*}

The cross-Kerr effect has been pointed out in the main text through Fig. \ref{fig:Fig7}, and it has been shown that
it enhances in-phase synchronization. This enhancement is related to the control of the 
frequency mismatch through $\tilde{\omega}_{1} =\omega_{1}-g_{ck}\alpha^{\ast}\alpha$.  As we have the frequency 
hierarchy of $\omega_{1}< \omega_{2}$, we have conveniently used a negative quadratic term 
($g_{\rm{ck}}<0$) in order to minimize the effect of $\delta\omega$. The dynamical 
states of the points labeled in Fig. \ref{fig:Fig7}b are shown in Fig. \ref{fig:Fig8}. It can be 
clearly seen that these dynamics agree well with the collective behaviours displayed in Fig. \ref{fig:Fig7}b. 
Furthermore, the dynamical state carried out in the gray area intertwined in 
the magenta zone in Fig. \ref{fig:Fig7}b is revealed. It results that, the mechanical resonators exhibit Rabi oscillations 
within this regime, which means that they are strongly coupled and can exchange energy.

\newpage

\bibliography{Synchronization}

\begin{thebibliography}{29}
\expandafter\ifx\csname natexlab\endcsname\relax\def\natexlab#1{#1}\fi
\expandafter\ifx\csname bibnamefont\endcsname\relax
  \def\bibnamefont#1{#1}\fi
\expandafter\ifx\csname bibfnamefont\endcsname\relax
  \def\bibfnamefont#1{#1}\fi
\expandafter\ifx\csname citenamefont\endcsname\relax
  \def\citenamefont#1{#1}\fi
\expandafter\ifx\csname url\endcsname\relax
  \def\url#1{\texttt{#1}}\fi
\expandafter\ifx\csname urlprefix\endcsname\relax\def\urlprefix{URL }\fi
\providecommand{\bibinfo}[2]{#2}
\providecommand{\eprint}[2][]{\url{#2}}

\bibitem[{\citenamefont{Eichenfield
  et~al.}(2009{\natexlab{a}})\citenamefont{Eichenfield, Chan, Camacho, Vahala,
  and Painter}}]{Eichenfield2009}
\bibinfo{author}{\bibfnamefont{M.}~\bibnamefont{Eichenfield}},
  \bibinfo{author}{\bibfnamefont{J.}~\bibnamefont{Chan}},
  \bibinfo{author}{\bibfnamefont{R.~M.} \bibnamefont{Camacho}},
  \bibinfo{author}{\bibfnamefont{K.~J.} \bibnamefont{Vahala}},
  \bibnamefont{and} \bibinfo{author}{\bibfnamefont{O.}~\bibnamefont{Painter}},
  \bibinfo{journal}{Nature} \textbf{\bibinfo{volume}{462}}, \bibinfo{pages}{78}
  (\bibinfo{year}{2009}{\natexlab{a}}).

\bibitem[{\citenamefont{Eichenfield
  et~al.}(2009{\natexlab{b}})\citenamefont{Eichenfield, Camacho, Chan, Vahala,
  and Painter}}]{Eichenfield_2009}
\bibinfo{author}{\bibfnamefont{M.}~\bibnamefont{Eichenfield}},
  \bibinfo{author}{\bibfnamefont{R.}~\bibnamefont{Camacho}},
  \bibinfo{author}{\bibfnamefont{J.}~\bibnamefont{Chan}},
  \bibinfo{author}{\bibfnamefont{K.~J.} \bibnamefont{Vahala}},
  \bibnamefont{and} \bibinfo{author}{\bibfnamefont{O.}~\bibnamefont{Painter}},
  \bibinfo{journal}{Nature} \textbf{\bibinfo{volume}{459}},
  \bibinfo{pages}{550} (\bibinfo{year}{2009}{\natexlab{b}}).

\bibitem[{\citenamefont{Chang et~al.}(2011)\citenamefont{Chang, Safavi-Naeini,
  Hafezi, and Painter}}]{Chang.2011.NJP}
\bibinfo{author}{\bibfnamefont{D.~E.} \bibnamefont{Chang}},
  \bibinfo{author}{\bibfnamefont{A.~H.} \bibnamefont{Safavi-Naeini}},
  \bibinfo{author}{\bibfnamefont{M.}~\bibnamefont{Hafezi}}, \bibnamefont{and}
  \bibinfo{author}{\bibfnamefont{O.}~\bibnamefont{Painter}},
  \bibinfo{journal}{New Journal of Physics} \textbf{\bibinfo{volume}{13}},
  \bibinfo{pages}{023003} (\bibinfo{year}{2011}).

\bibitem[{\citenamefont{Heinrich et~al.}(2011)\citenamefont{Heinrich, Ludwig,
  Qian, Kubala, and Marquardt}}]{Heinrich.2011}
\bibinfo{author}{\bibfnamefont{G.}~\bibnamefont{Heinrich}},
  \bibinfo{author}{\bibfnamefont{M.}~\bibnamefont{Ludwig}},
  \bibinfo{author}{\bibfnamefont{J.}~\bibnamefont{Qian}},
  \bibinfo{author}{\bibfnamefont{B.}~\bibnamefont{Kubala}}, \bibnamefont{and}
  \bibinfo{author}{\bibfnamefont{F.}~\bibnamefont{Marquardt}},
  \bibinfo{journal}{Phys. Rev. Lett.} \textbf{\bibinfo{volume}{107}},
  \bibinfo{pages}{043603} (\bibinfo{year}{2011}),
  \urlprefix\url{https://link.aps.org/doi/10.1103/PhysRevLett.107.043603}.

\bibitem[{\citenamefont{Wallin et~al.}(2018)\citenamefont{Wallin, De~Alba,
  Westly, Holland, Grutzik, Rand, Zehnder, Aksyuk, Krylov, and
  Ilic}}]{Wallin.2018}
\bibinfo{author}{\bibfnamefont{C.~B.} \bibnamefont{Wallin}},
  \bibinfo{author}{\bibfnamefont{R.}~\bibnamefont{De~Alba}},
  \bibinfo{author}{\bibfnamefont{D.}~\bibnamefont{Westly}},
  \bibinfo{author}{\bibfnamefont{G.}~\bibnamefont{Holland}},
  \bibinfo{author}{\bibfnamefont{S.}~\bibnamefont{Grutzik}},
  \bibinfo{author}{\bibfnamefont{R.~H.} \bibnamefont{Rand}},
  \bibinfo{author}{\bibfnamefont{A.~T.} \bibnamefont{Zehnder}},
  \bibinfo{author}{\bibfnamefont{V.~A.} \bibnamefont{Aksyuk}},
  \bibinfo{author}{\bibfnamefont{S.}~\bibnamefont{Krylov}}, \bibnamefont{and}
  \bibinfo{author}{\bibfnamefont{B.~R.} \bibnamefont{Ilic}},
  \bibinfo{journal}{Phys. Rev. Lett.} \textbf{\bibinfo{volume}{121}},
  \bibinfo{pages}{264301} (\bibinfo{year}{2018}),
  \urlprefix\url{https://link.aps.org/doi/10.1103/PhysRevLett.121.264301}.

\bibitem[{\citenamefont{Ludwig and Marquardt}(2013)}]{Ludwig.2013.Sync}
\bibinfo{author}{\bibfnamefont{M.}~\bibnamefont{Ludwig}} \bibnamefont{and}
  \bibinfo{author}{\bibfnamefont{F.}~\bibnamefont{Marquardt}},
  \bibinfo{journal}{Phys. Rev. Lett.} \textbf{\bibinfo{volume}{111}},
  \bibinfo{pages}{073603} (\bibinfo{year}{2013}),
  \urlprefix\url{https://link.aps.org/doi/10.1103/PhysRevLett.111.073603}.

\bibitem[{\citenamefont{Xuereb et~al.}(2012)\citenamefont{Xuereb, Genes, and
  Dantan}}]{Xuereb.2012}
\bibinfo{author}{\bibfnamefont{A.}~\bibnamefont{Xuereb}},
  \bibinfo{author}{\bibfnamefont{C.}~\bibnamefont{Genes}}, \bibnamefont{and}
  \bibinfo{author}{\bibfnamefont{A.}~\bibnamefont{Dantan}},
  \bibinfo{journal}{Phys. Rev. Lett.} \textbf{\bibinfo{volume}{109}},
  \bibinfo{pages}{223601} (\bibinfo{year}{2012}),
  \urlprefix\url{https://link.aps.org/doi/10.1103/PhysRevLett.109.223601}.

\bibitem[{\citenamefont{Xuereb et~al.}(2014)\citenamefont{Xuereb, Genes,
  Pupillo, Paternostro, and Dantan}}]{Xuereb.2014}
\bibinfo{author}{\bibfnamefont{A.}~\bibnamefont{Xuereb}},
  \bibinfo{author}{\bibfnamefont{C.}~\bibnamefont{Genes}},
  \bibinfo{author}{\bibfnamefont{G.}~\bibnamefont{Pupillo}},
  \bibinfo{author}{\bibfnamefont{M.}~\bibnamefont{Paternostro}},
  \bibnamefont{and} \bibinfo{author}{\bibfnamefont{A.}~\bibnamefont{Dantan}},
  \bibinfo{journal}{Phys. Rev. Lett.} \textbf{\bibinfo{volume}{112}},
  \bibinfo{pages}{133604} (\bibinfo{year}{2014}),
  \urlprefix\url{https://link.aps.org/doi/10.1103/PhysRevLett.112.133604}.

\bibitem[{\citenamefont{Chen and Clerk}(2014)}]{Chen.2014}
\bibinfo{author}{\bibfnamefont{W.}~\bibnamefont{Chen}} \bibnamefont{and}
  \bibinfo{author}{\bibfnamefont{A.~A.} \bibnamefont{Clerk}},
  \bibinfo{journal}{Phys. Rev. A} \textbf{\bibinfo{volume}{89}},
  \bibinfo{pages}{033854} (\bibinfo{year}{2014}),
  \urlprefix\url{https://link.aps.org/doi/10.1103/PhysRevA.89.033854}.

\bibitem[{\citenamefont{Schmidt et~al.}(2015)\citenamefont{Schmidt, Peano, and
  Marquardt}}]{Schmidt.2015}
\bibinfo{author}{\bibfnamefont{M.}~\bibnamefont{Schmidt}},
  \bibinfo{author}{\bibfnamefont{V.}~\bibnamefont{Peano}}, \bibnamefont{and}
  \bibinfo{author}{\bibfnamefont{F.}~\bibnamefont{Marquardt}},
  \bibinfo{journal}{New Journal of Physics} \textbf{\bibinfo{volume}{17}},
  \bibinfo{pages}{023025} (\bibinfo{year}{2015}).

\bibitem[{\citenamefont{Roque et~al.}(2017)\citenamefont{Roque, Peano,
  Yevtushenko, and Marquardt}}]{Roque.2017}
\bibinfo{author}{\bibfnamefont{T.~F.} \bibnamefont{Roque}},
  \bibinfo{author}{\bibfnamefont{V.}~\bibnamefont{Peano}},
  \bibinfo{author}{\bibfnamefont{O.~M.} \bibnamefont{Yevtushenko}},
  \bibnamefont{and}
  \bibinfo{author}{\bibfnamefont{F.}~\bibnamefont{Marquardt}},
  \bibinfo{journal}{New Journal of Physics} \textbf{\bibinfo{volume}{19}},
  \bibinfo{pages}{013006} (\bibinfo{year}{2017}).

\bibitem[{\citenamefont{Peano et~al.}(2015)\citenamefont{Peano, Brendel,
  Schmidt, and Marquardt}}]{Peano.2015.PRL}
\bibinfo{author}{\bibfnamefont{V.}~\bibnamefont{Peano}},
  \bibinfo{author}{\bibfnamefont{C.}~\bibnamefont{Brendel}},
  \bibinfo{author}{\bibfnamefont{M.}~\bibnamefont{Schmidt}}, \bibnamefont{and}
  \bibinfo{author}{\bibfnamefont{F.}~\bibnamefont{Marquardt}},
  \bibinfo{journal}{Phys. Rev. X} \textbf{\bibinfo{volume}{5}},
  \bibinfo{pages}{031011} (\bibinfo{year}{2015}),
  \urlprefix\url{https://link.aps.org/doi/10.1103/PhysRevX.5.031011}.

\bibitem[{\citenamefont{Brendel et~al.}(2018)\citenamefont{Brendel, Peano,
  Painter, and Marquardt}}]{Brendel.2018}
\bibinfo{author}{\bibfnamefont{C.}~\bibnamefont{Brendel}},
  \bibinfo{author}{\bibfnamefont{V.}~\bibnamefont{Peano}},
  \bibinfo{author}{\bibfnamefont{O.}~\bibnamefont{Painter}}, \bibnamefont{and}
  \bibinfo{author}{\bibfnamefont{F.}~\bibnamefont{Marquardt}},
  \bibinfo{journal}{Phys. Rev. B} \textbf{\bibinfo{volume}{97}},
  \bibinfo{pages}{020102} (\bibinfo{year}{2018}),
  \urlprefix\url{https://link.aps.org/doi/10.1103/PhysRevB.97.020102}.

\bibitem[{\citenamefont{Bregni}(2002)}]{Bregni.2002}
\bibinfo{author}{\bibfnamefont{S.}~\bibnamefont{Bregni}},
  \emph{\bibinfo{title}{Synchronization of Digital Telecommunications
  Networks}} (\bibinfo{publisher}{Wiley}, \bibinfo{year}{2002}).

\bibitem[{\citenamefont{Strogatz}(2003)}]{strogatz2003}
\bibinfo{author}{\bibfnamefont{S.}~\bibnamefont{Strogatz}},
  \emph{\bibinfo{title}{Sync : the emerging science of spontaneous order}}
  (\bibinfo{publisher}{Hyperion}, \bibinfo{address}{New York},
  \bibinfo{year}{2003}), ISBN \bibinfo{isbn}{0-7868-6844-9}.

\bibitem[{\citenamefont{Bahder}(2009)}]{bahder2009}
\bibinfo{author}{\bibfnamefont{T.}~\bibnamefont{Bahder}},
  \emph{\bibinfo{title}{Clock synchronization and navigation in the vicinity of
  the earth}} (\bibinfo{publisher}{Nova Science}, \bibinfo{address}{New York},
  \bibinfo{year}{2009}), ISBN \bibinfo{isbn}{978-1606921142}.

\bibitem[{\citenamefont{Zhang et~al.}(2012)\citenamefont{Zhang, Wiederhecker,
  Manipatruni, Barnard, McEuen, and Lipson}}]{Zhang.2012}
\bibinfo{author}{\bibfnamefont{M.}~\bibnamefont{Zhang}},
  \bibinfo{author}{\bibfnamefont{G.~S.} \bibnamefont{Wiederhecker}},
  \bibinfo{author}{\bibfnamefont{S.}~\bibnamefont{Manipatruni}},
  \bibinfo{author}{\bibfnamefont{A.}~\bibnamefont{Barnard}},
  \bibinfo{author}{\bibfnamefont{P.}~\bibnamefont{McEuen}}, \bibnamefont{and}
  \bibinfo{author}{\bibfnamefont{M.}~\bibnamefont{Lipson}},
  \bibinfo{journal}{Phys. Rev. Lett.} \textbf{\bibinfo{volume}{109}},
  \bibinfo{pages}{233906} (\bibinfo{year}{2012}),
  \urlprefix\url{https://link.aps.org/doi/10.1103/PhysRevLett.109.233906}.

\bibitem[{\citenamefont{Zhang et~al.}(2015)\citenamefont{Zhang, Shah, Cardenas,
  and Lipson}}]{Zhang_2015}
\bibinfo{author}{\bibfnamefont{M.}~\bibnamefont{Zhang}},
  \bibinfo{author}{\bibfnamefont{S.}~\bibnamefont{Shah}},
  \bibinfo{author}{\bibfnamefont{J.}~\bibnamefont{Cardenas}}, \bibnamefont{and}
  \bibinfo{author}{\bibfnamefont{M.}~\bibnamefont{Lipson}},
  \bibinfo{journal}{Phys. Rev. Lett.} \textbf{\bibinfo{volume}{115}},
  \bibinfo{pages}{163902} (\bibinfo{year}{2015}),
  \urlprefix\url{https://link.aps.org/doi/10.1103/PhysRevLett.115.163902}.

\bibitem[{\citenamefont{Bagheri et~al.}(2013)\citenamefont{Bagheri, Poot, Fan,
  Marquardt, and Tang}}]{Bagheri.2013}
\bibinfo{author}{\bibfnamefont{M.}~\bibnamefont{Bagheri}},
  \bibinfo{author}{\bibfnamefont{M.}~\bibnamefont{Poot}},
  \bibinfo{author}{\bibfnamefont{L.}~\bibnamefont{Fan}},
  \bibinfo{author}{\bibfnamefont{F.}~\bibnamefont{Marquardt}},
  \bibnamefont{and} \bibinfo{author}{\bibfnamefont{H.~X.} \bibnamefont{Tang}},
  \bibinfo{journal}{Phys. Rev. Lett.} \textbf{\bibinfo{volume}{111}},
  \bibinfo{pages}{213902} (\bibinfo{year}{2013}),
  \urlprefix\url{https://link.aps.org/doi/10.1103/PhysRevLett.111.213902}.

\bibitem[{\citenamefont{Colombano et~al.}(2019)\citenamefont{Colombano,
  Arregui, Capuj, Pitanti, Maire, Griol, Garrido, Martinez, Sotomayor-Torres,
  and Navarro-Urrios}}]{Colombano.2019}
\bibinfo{author}{\bibfnamefont{M.~F.} \bibnamefont{Colombano}},
  \bibinfo{author}{\bibfnamefont{G.}~\bibnamefont{Arregui}},
  \bibinfo{author}{\bibfnamefont{N.~E.} \bibnamefont{Capuj}},
  \bibinfo{author}{\bibfnamefont{A.}~\bibnamefont{Pitanti}},
  \bibinfo{author}{\bibfnamefont{J.}~\bibnamefont{Maire}},
  \bibinfo{author}{\bibfnamefont{A.}~\bibnamefont{Griol}},
  \bibinfo{author}{\bibfnamefont{B.}~\bibnamefont{Garrido}},
  \bibinfo{author}{\bibfnamefont{A.}~\bibnamefont{Martinez}},
  \bibinfo{author}{\bibfnamefont{C.~M.} \bibnamefont{Sotomayor-Torres}},
  \bibnamefont{and}
  \bibinfo{author}{\bibfnamefont{D.}~\bibnamefont{Navarro-Urrios}},
  \bibinfo{journal}{Phys. Rev. Lett.} \textbf{\bibinfo{volume}{123}},
  \bibinfo{pages}{017402} (\bibinfo{year}{2019}),
  \urlprefix\url{https://link.aps.org/doi/10.1103/PhysRevLett.123.017402}.

\bibitem[{\citenamefont{Sheng et~al.}(2020)\citenamefont{Sheng, Wei, Yang, and
  Wu}}]{Sheng.2020}
\bibinfo{author}{\bibfnamefont{J.}~\bibnamefont{Sheng}},
  \bibinfo{author}{\bibfnamefont{X.}~\bibnamefont{Wei}},
  \bibinfo{author}{\bibfnamefont{C.}~\bibnamefont{Yang}}, \bibnamefont{and}
  \bibinfo{author}{\bibfnamefont{H.}~\bibnamefont{Wu}}, \bibinfo{journal}{Phys.
  Rev. Lett.} \textbf{\bibinfo{volume}{124}}, \bibinfo{pages}{053604}
  (\bibinfo{year}{2020}),
  \urlprefix\url{https://link.aps.org/doi/10.1103/PhysRevLett.124.053604}.

\bibitem[{\citenamefont{Shah et~al.}(2015)\citenamefont{Shah, Zhang, Rand, and
  Lipson}}]{Shah.2015}
\bibinfo{author}{\bibfnamefont{S.~Y.} \bibnamefont{Shah}},
  \bibinfo{author}{\bibfnamefont{M.}~\bibnamefont{Zhang}},
  \bibinfo{author}{\bibfnamefont{R.}~\bibnamefont{Rand}}, \bibnamefont{and}
  \bibinfo{author}{\bibfnamefont{M.}~\bibnamefont{Lipson}},
  \bibinfo{journal}{Phys. Rev. Lett.} \textbf{\bibinfo{volume}{114}},
  \bibinfo{pages}{113602} (\bibinfo{year}{2015}),
  \urlprefix\url{https://link.aps.org/doi/10.1103/PhysRevLett.114.113602}.

\bibitem[{\citenamefont{Gil-Santos et~al.}(2017)\citenamefont{Gil-Santos,
  Labousse, Baker, Goetschy, Hease, Gomez, Lema\^{\i}tre, Leo, Ciuti, and
  Favero}}]{Gil.2017}
\bibinfo{author}{\bibfnamefont{E.}~\bibnamefont{Gil-Santos}},
  \bibinfo{author}{\bibfnamefont{M.}~\bibnamefont{Labousse}},
  \bibinfo{author}{\bibfnamefont{C.}~\bibnamefont{Baker}},
  \bibinfo{author}{\bibfnamefont{A.}~\bibnamefont{Goetschy}},
  \bibinfo{author}{\bibfnamefont{W.}~\bibnamefont{Hease}},
  \bibinfo{author}{\bibfnamefont{C.}~\bibnamefont{Gomez}},
  \bibinfo{author}{\bibfnamefont{A.}~\bibnamefont{Lema\^{\i}tre}},
  \bibinfo{author}{\bibfnamefont{G.}~\bibnamefont{Leo}},
  \bibinfo{author}{\bibfnamefont{C.}~\bibnamefont{Ciuti}}, \bibnamefont{and}
  \bibinfo{author}{\bibfnamefont{I.}~\bibnamefont{Favero}},
  \bibinfo{journal}{Phys. Rev. Lett.} \textbf{\bibinfo{volume}{118}},
  \bibinfo{pages}{063605} (\bibinfo{year}{2017}),
  \urlprefix\url{https://link.aps.org/doi/10.1103/PhysRevLett.118.063605}.

\bibitem[{\citenamefont{Hong et~al.}(2017)\citenamefont{Hong, Riedinger,
  Marinkovi{\'{c}}, Wallucks, Hofer, Norte, Aspelmeyer, and
  Gröblacher}}]{Hong.2017}
\bibinfo{author}{\bibfnamefont{S.}~\bibnamefont{Hong}},
  \bibinfo{author}{\bibfnamefont{R.}~\bibnamefont{Riedinger}},
  \bibinfo{author}{\bibfnamefont{I.}~\bibnamefont{Marinkovi{\'{c}}}},
  \bibinfo{author}{\bibfnamefont{A.}~\bibnamefont{Wallucks}},
  \bibinfo{author}{\bibfnamefont{S.~G.} \bibnamefont{Hofer}},
  \bibinfo{author}{\bibfnamefont{R.~A.} \bibnamefont{Norte}},
  \bibinfo{author}{\bibfnamefont{M.}~\bibnamefont{Aspelmeyer}},
  \bibnamefont{and}
  \bibinfo{author}{\bibfnamefont{S.}~\bibnamefont{Gröblacher}},
  \bibinfo{journal}{Science} \textbf{\bibinfo{volume}{358}},
  \bibinfo{pages}{203} (\bibinfo{year}{2017}).

\bibitem[{\citenamefont{Djorwe et~al.}(2018)\citenamefont{Djorwe, Pennec, and
  Djafari-Rouhani}}]{Djorwe.2018.PRE}
\bibinfo{author}{\bibfnamefont{P.}~\bibnamefont{Djorwe}},
  \bibinfo{author}{\bibfnamefont{Y.}~\bibnamefont{Pennec}}, \bibnamefont{and}
  \bibinfo{author}{\bibfnamefont{B.}~\bibnamefont{Djafari-Rouhani}},
  \bibinfo{journal}{Phys. Rev. E} \textbf{\bibinfo{volume}{98}},
  \bibinfo{pages}{032201} (\bibinfo{year}{2018}),
  \urlprefix\url{https://link.aps.org/doi/10.1103/PhysRevE.98.032201}.

\bibitem[{\citenamefont{Karnatak et~al.}(2010)\citenamefont{Karnatak, Punetha,
  Prasad, and Ramaswamy}}]{Karnatak.2010}
\bibinfo{author}{\bibfnamefont{R.}~\bibnamefont{Karnatak}},
  \bibinfo{author}{\bibfnamefont{N.}~\bibnamefont{Punetha}},
  \bibinfo{author}{\bibfnamefont{A.}~\bibnamefont{Prasad}}, \bibnamefont{and}
  \bibinfo{author}{\bibfnamefont{R.}~\bibnamefont{Ramaswamy}},
  \bibinfo{journal}{Phys. Rev. E} \textbf{\bibinfo{volume}{82}},
  \bibinfo{pages}{046219} (\bibinfo{year}{2010}),
  \urlprefix\url{https://link.aps.org/doi/10.1103/PhysRevE.82.046219}.

\bibitem[{\citenamefont{Sharma et~al.}(2012)\citenamefont{Sharma, Shrimali, and
  Dana}}]{Sharma.2012}
\bibinfo{author}{\bibfnamefont{A.}~\bibnamefont{Sharma}},
  \bibinfo{author}{\bibfnamefont{M.~D.} \bibnamefont{Shrimali}},
  \bibnamefont{and} \bibinfo{author}{\bibfnamefont{S.~K.} \bibnamefont{Dana}},
  \bibinfo{journal}{Chaos: An Interdisciplinary Journal of Nonlinear Science}
  \textbf{\bibinfo{volume}{22}}, \bibinfo{pages}{023147}
  (\bibinfo{year}{2012}).

\bibitem[{\citenamefont{Djorwe et~al.}(2019)\citenamefont{Djorwe, Pennec, and
  Djafari-Rouhani}}]{Djorwe.2019.sensor}
\bibinfo{author}{\bibfnamefont{P.}~\bibnamefont{Djorwe}},
  \bibinfo{author}{\bibfnamefont{Y.}~\bibnamefont{Pennec}}, \bibnamefont{and}
  \bibinfo{author}{\bibfnamefont{B.}~\bibnamefont{Djafari-Rouhani}},
  \bibinfo{journal}{Phys. Rev. Applied} \textbf{\bibinfo{volume}{12}},
  \bibinfo{pages}{024002} (\bibinfo{year}{2019}),
  \urlprefix\url{https://link.aps.org/doi/10.1103/PhysRevApplied.12.024002}.

\bibitem[{\citenamefont{Para\"{\i}so et~al.}(2015)\citenamefont{Para\"{\i}so,
  Kalaee, Zang, Pfeifer, Marquardt, and Painter}}]{Paraiso.2015}
\bibinfo{author}{\bibfnamefont{T.~K.} \bibnamefont{Para\"{\i}so}},
  \bibinfo{author}{\bibfnamefont{M.}~\bibnamefont{Kalaee}},
  \bibinfo{author}{\bibfnamefont{L.}~\bibnamefont{Zang}},
  \bibinfo{author}{\bibfnamefont{H.}~\bibnamefont{Pfeifer}},
  \bibinfo{author}{\bibfnamefont{F.}~\bibnamefont{Marquardt}},
  \bibnamefont{and} \bibinfo{author}{\bibfnamefont{O.}~\bibnamefont{Painter}},
  \bibinfo{journal}{Phys. Rev. X} \textbf{\bibinfo{volume}{5}},
  \bibinfo{pages}{041024} (\bibinfo{year}{2015}),
  \urlprefix\url{https://link.aps.org/doi/10.1103/PhysRevX.5.041024}.

\end{thebibliography}

\end{document}